\documentclass[aps,prl,twocolumn,superscriptaddress,groupedaddress]{revtex4}  
\usepackage{graphicx}  
\usepackage{dcolumn}   
\usepackage{bm}        
\usepackage{amssymb}   

\begin{document}



\title{Constraints on neutrino masses from 
Baryon Acoustic Oscillation measurements}
\author{B.~Hoeneisen} \affiliation{Universidad San Francisco de Quito, Quito, Ecuador}
\date{December 10, 2017}

\begin{abstract}
From 21 independent Baryon Acoustic Oscillation (BAO) measurements
we obtain the following sum of masses of 
active Dirac or Majorana neutrinos:
$\sum m_\nu = 0.711 - 0.335 \cdot \delta h + 0.050 \cdot \delta b
        \pm 0.063 \textrm{ eV,}$
where
$\delta h \equiv (h - 0.678) / 0.009$ and
$\delta b \equiv (\Omega_b h^2 - 0.02226) / 0.00023$.
This result may be combined with independent measurements
that constrain the parameters $\sum m_\nu$, $h$, and $\Omega_b h^2$.
For $\delta h = \pm 1$ and $\delta b = \pm 1$, we obtain
$m_\nu < 0.43$ eV at 95\% confidence.
\end{abstract}

\maketitle


We extend the analysis presented in Ref. \cite{ijaa} to include
neutrino masses. The present analysis has three steps:
(1) we calculate the distance of propagation $r_s$, in units of
$c/H_0$, referred to the present time, of sound waves
in the photon-electron-baryon plasma until
decoupling by numerical integration of Eqs. (16) and (17) of
Ref. \cite{ijaa}; 
(2) we fit the Friedmann equation of evolution of
the universe to
21 independent Baryon Acoustic
Oscillation (BAO) distance measurements listed in \cite{ijaa}
used as uncalibrated standard rulers
and obtain the length $d$ of these rulers, in
units of $c/H_0$, referred to the present time; and
(3) we set
\begin{equation}
r_s = d
\label{rs_d}
\end{equation}
to constrain the sum of neutrino masses $\sum m_\nu$.
$c$ is the speed of light, and 
$H_0 \equiv 100 h$ km s$^{-1}$ Mpc$^{-1}$ is the present day
Hubble expansion parameter. 

The main body of this article assumes:
(1) flat space, i.e. $\Omega_k = 0$, and
(2) constant dark energy density relative to the critical density,
i.e. $\Omega_\textrm{DE}$ independent of the expansion parameter
$a$. These constraints
are in agreement with all observations to date \cite{ijaa, PDG2016}.
Results without these constraints are presented in the appendix.

To be specific we consider three active neutrino flavors with 
three eigenstates with nearly the same 
mass $m_\nu$, so $\sum m_\nu = 3 m_\nu$.
This is a useful scenario to consider since our current
limits on $m_\nu^2$ are much larger than the 
mass-squared-differences $\Delta m^2$ and $\Delta m^2_{21}$
obtained from neutrino oscillations \cite{PDG2016}.
These neutrinos become
non-relativistic at a neutrino temperature
$T_\nu = m_\nu / 3.15$ or a photon temperature
$T = m_\nu (11/4)^{1/3} / 3.15$. The corresponding
expansion parameter is 
$a_\nu = T_0/T = 5.28 \times 10^{-4} (1 eV / m_\nu)$. 

The matter density relative to the present critical density 
is $\Omega_m / a^3$ for $a > a_\nu$.
$\Omega_m$ includes the density 
$\Omega_\nu = h^{-2} \sum m_\nu / 94 \textrm{eV}$ of
Dirac or Majorana neutrinos that are non-relativistic today.
Note that for Dirac neutrinos we are considering the
scenario in which right-handed neutrinos and left-handed
anti-neutrinos are sterile and never achieved
thermal equilibrium. Our results can be amended for
other specific scenarios.
For $a < a_\nu$ we take the matter density to be
$(\Omega_m - \Omega_\nu) / a^3$. 
The radiation density is 
$\Omega_\gamma N_\textrm{eq} / (2 a^4)$
for $a < a_\nu$, where 
$N_\textrm{eq} = 3.36$ for three flavors of Dirac
(mostly) left-handed neutrinos and right-handed anti-neutrinos.
We also take $N_\textrm{eq} = 3.36$ for three active flavors of
Majorana left-handed and right-handed neutrinos.
For $a > a_\nu$, we take the
radiation density to be 
$( \Omega_\gamma N_\textrm{eq} / 2 - a_\nu \Omega_\nu ) / a^4 = 
\Omega_\gamma / a^4$.
The present density of photons relative to the critical density is
$\Omega_\gamma = 2.473 \times 10^{-5}  h^{-2}$ \cite{PDG2016}.

The data used to obtain $d$ are 18 independent BAO
distance measurements with Sloan Digital Sky Survey (SDSS) data release
DR13 galaxies in the redshift range $z = 0.1$ to 0.7 \cite{ijaa, DR13, BOSS, eBOSS}, 
two BAO distance measurements in the Lyman-alpha forest 
(Ly$\alpha$) at $z = 2.36$ (cross-correlation \cite{lyman}) and $z = 2.34$ 
(autocorrelation \cite{lyman2}), 
and the Cosmic Microwave Background (CMB) correlation angle 
$\theta_\textrm{MC} = 0.010410 \pm 0.000005$ \cite{PDG2016, Planck}, 
used as an uncalibrated standard ruler.
These 21 independent BAO measurements are summarized in \cite{ijaa}.

As a reference we take 
\begin{equation}
h = 0.678 \pm 0.009, \qquad \Omega_b h^2 = 0.02226 \pm 0.00023
\end{equation}
(at 68\% confidence) from ``Planck TT $+$ low P $+$ lensing"
data (that does not contain BAO information) \cite{PDG2016}.
$\Omega_b$ is the present density of baryons
relative to the critical density.

Due to correlations and non-linearities we 
obtain our final result (Eq. (\ref{sum2}) below) 
with a global fit. The following equations are included to
illustrate the dependence of $r_s$ and $d$ on the cosmological
parameters $h$, $\Omega_b h^2$ and $\sum m_\nu$ in limited ranges 
of interest. 
Integrating the comoving sound speed of the photon-baryon-electron plasma
until $a_\textrm{dec} = 1/(1 + z_\textrm{dec})$ with 
$z_\textrm{dec} = 1089.9 \pm 0.4$ \cite{PDG2016} we obtain
\begin{equation}
r_s \approx 0.0339 \times A \times \left( \frac{0.28}{\Omega_m} \right)^{0.24}
\label{rs}
\end{equation}
with 
\begin{equation}
A \approx 0.990 + 0.007 \cdot \delta h - 0.001 \cdot \delta b + 0.020 \cdot \sum m_\nu, 
\label{rs2}
\end{equation}
where
\begin{eqnarray}
\delta h & \equiv & (h - 0.678) / 0.009, \\ 
\delta b & \equiv & (\Omega_b h^2 - 0.02226) / 0.00023.
\end{eqnarray}

To obtain $d$ we minimize the $\chi^2$ with 21 terms, corresponding to the
21 BAO observables, with respect to $\Omega_\textrm{DE}$ and $d$,
and obtain $O_\textrm{DE} = 0.718 \pm 0.003$ and 
\begin{equation}
d \approx 0.0340 \pm 0.0002,
\end{equation}
with $\chi^2$ per degree of freedom $19.8/19$, and
correlation coefficient 0.989 (this high correlation
coefficient is due to the high precision of $\theta_\textrm{MC}$).
Setting $r_s = d$ we obtain 
\begin{equation}
\sum m_\nu \approx 0.73 - 0.35 \cdot \delta h + 0.05 \cdot \delta b 
        \pm 0.15 \textrm{ eV.}
\label{sum}
\end{equation}

A more precise result is obtained with a global fit by
minimizing the $\chi^2$ with 21 terms varying
$\Omega_\textrm{DE}$ and $\sum m_\nu$ directly. We obtain 
$\Omega_\textrm{DE} = 0.7175 \pm 0.0023$ and
\begin{equation}
\sum m_\nu = 0.711 - 0.335 \cdot \delta h + 0.050 \cdot \delta b
        \pm 0.063 \textrm{ eV,}
\label{sum2}
\end{equation}
with $\chi^2 / \textrm{d.f.} = 19.9/19$, and correlation
coefficient 0.924.
This is our main result. Equation (\ref{sum2}) is obtained
from BAO measurements alone, and
is written in a way
that can be combined with independent constraints on the
cosmological parameters $\sum m_\nu$, $h$ and $\Omega_b h^2$,
such as measurements of the power spectrum of density
fluctuations $P(k)$, the CMB, and direct measurements
of the Hubble parameter.

Setting $\delta h = \pm 1$ and $\delta b = \pm 1$ we obtain the
following upper bound on the mass of active neutrinos 
$m_\nu = \frac{1}{3} \sum m_\nu$:
\begin{equation}
m_\nu < 0.43 \textrm{ eV at 95\% confidence}.
\end{equation}

\appendix

\section{Appendix}

Freeing $\Omega_k$ and keeping $\Omega_\textrm{DE}$ constant
we obtain 
$\Omega_k = -0.003 \pm 0.006$,
$\Omega_\textrm{DE} + 2.2 \Omega_k = 0.719 \pm 0.003$, and
\begin{equation}
\sum m_\nu = 0.623 - 0.334 \cdot \delta h + 0.050 \cdot \delta b 
     \pm 0.191 \textrm{ eV,}
\end{equation}
with $\chi^2/\textrm{d.f.} = 19.6/18$.

Fixing $\Omega_k = 0$ and letting 
$\Omega_\textrm{DE}(a) = \Omega_\textrm{DE} \cdot \{1 + w_a \cdot (1 - a)\}$
we obtain 
$\Omega_\textrm{DE} = 0.716 \pm 0.004$,
$w_a = 0.064 \pm 0.148$, and
\begin{equation}
\sum m_\nu = 0.603 - 0.349 \cdot \delta h + 0.052 \cdot \delta b 
     \pm 0.257 \textrm{ eV,}
\end{equation}
with $\chi^2/\textrm{d.f.} = 19.7/18$.

Freeing $\Omega_k$ and letting 
$\Omega_\textrm{DE}(a) = \Omega_\textrm{DE} \cdot \{1 + w_a \cdot (1 - a)\}$
we obtain 
$\Omega_k = -0.008 \pm 0.004$,
$\Omega_\textrm{DE} + 2.2 \Omega_k = 0.718 \pm 0.004$,
$w_a = 0.227 \pm 0.069$, and
\begin{equation}
0 < \sum m_\nu = -0.388 - 0.350 \cdot \delta h + 0.050 \cdot \delta b 
     \pm 0.830 \textrm{ eV,}
\end{equation}
with $\chi^2/\textrm{d.f.} = 17.8/17$.

Full details of the fitting method are presented in \cite{ijaa}.

\end{document}